\documentclass[aps,prl,twocolumn,nofootinbib,superscriptaddress,amssymb]{revtex4}
\usepackage{color}
\usepackage{epsfig}
\usepackage{graphicx}
\usepackage{epstopdf}
\usepackage{mathtools}
\usepackage{bbold}
\def\hhh{\Delta}

\def\be{\begin{equation}}
\def\ee{\end{equation}}

\def\la{\langle}
\def\bea{\begin{eqnarray}}
\def\eea{\end{eqnarray}}
\def\is{\! & \! = \! & \!}
\def\ra{\rangle}

\def\ba{\begin{eqnarray}}
\def\ea{\end{eqnarray}}
\def\be{\bea}
\def\ee{\eea}

\newcommand{\smpc}{\hspace{.5pt}}
\def\ra{\bigr\rangle}
\def\la{\bigl\langle}

\def\nspc{\!\spc\smpc}

\def\ra{\bigr\rangle}
\def\la{\bigl\langle}
\def\li{\bigl|\spc}
\def\ri{\bigr |\spc}

\def\spc{\hspace{1pt}}
\def\nmpc{\hspace{-.5pt}}
\def\nspc{\hspace{-1pt}}
\def\smpc{\hspace{.5pt}}
\def\li{\bigl |}
\def\ra{\bigr \rangle}
\def\la{\bigl\langle}
\def\aa{\mbox{\small $A$}}
\def\bb{\mbox{\small $B$}}
\def\cc{\mbox{\small $C$}}
\def\hh{h}
\begin{document}

\setcounter{tocdepth}{2}
%\tableofcontents
\addtolength{\baselineskip}{.15mm} 
\addtolength{\parskip}{.4mm}
\addtolength{\abovedisplayskip}{.5mm}
\addtolength{\belowdisplayskip}{.5mm}

\title{Poking Holes in AdS/CFT:\\[1.5mm]
Bulk Fields from Boundary States %\\[2mm] 
}

\author{Herman Verlinde}
\affiliation{Department of Physics, Princeton University, Princeton, NJ 08544, USA}
\affiliation{Princeton Center for Theoretical Science, Princeton University, Princeton, NJ 08544, USA}

\date{\today}

\begin{abstract}
We propose an intrinsic CFT definition of local bulk operators in AdS${}_3$/CFT${}_{2}$ in terms of twisted Ishibashi boundary states. The bulk field $\Phi(X)$ creates a cross cap, a circular hole with opposite edge points identified, in the CFT space-time. The size of the hole is parameterized by the holographic radial coordinate $y$. Our definition is state-independent, non-perturbative, and does not presume or utilize a semi-classical bulk geometry. We argue that, at large $c$, the matrix element %$\langle \Delta+\omega | \Phi(X)| \Delta\rangle$ 
between highly excited states %with $\Delta\gg \frac{c}{12}\gg\omega$ 
satisfies the bulk wave equation in the AdS black hole background. 
%With our definition, the black hole space-time acquires a semi-classical meaning in the CFT as the moduli space of the cross cap. defect created by the bulk field. 
\end{abstract}

\def\calO{{b}}
\def\be{\begin{equation}}
\def\ee{\end{equation}}

%\begin{document}

% insert suggested PACS numbers in braces on next line
% insert suggested keywords - APS authors don't need to do this
%\keywords{}

%\maketitle must follow title, authors, abstract, \pacs, and \keywords
\maketitle
\vspace{-2mm}
\begin{center}
{ \bf Introduction}
\end{center}
\vspace{-3mm}

AdS/CFT duality has passed many tests. Most  checks compare CFT correlation functions with the dependence of AdS quantities on sources at the boundary \cite{adscft}. The CFT construction of local bulk observables  \cite{bulk1,bulk2},  on the other hand, remains underdeveloped. In this note we propose a new representation of local bulk fields in terms of operators that create  finite size holes in the space-time of the CFT. We formulate and test our proposal for the case of AdS${}_3$/CFT${}_2$.  
In the following, $x = (z,\bar{z})$ and $X= (y,x)$ denote coordinate systems on the 2D  space-time and in AdS${}_3$, respectively.  

Gauge/gravity duality postulates a one-to one map between the Hilbert space of the CFT and the gravity theory. Local bulk  fields should thus have images as suitable non-local operators in the CFT.
In particular, if the gravity side is weakly coupled, we can associate  to every local CFT operator ${\cal O}_h(x)$ an effective  field
$\Phi_h(X)$ in AdS, such that, as we approach the AdS boundary
\bea
\label{phitoo}
\lim_{y\to 0}\, y^{-2h} \Phi_\hh(y,x) = {\cal O}_h(x).
\eea
To leading order in $1/N$, the linearized bulk field satisfies a free field wave equation
\bea
\label{waveeq}
\square_{{}_{\rm bulk}}\nspc \Phi_\hh = m_\hh^2 \spc \Phi_\hh
\eea
with $m_\hh^2 =  \hh(\hh-d)$. This fact suggests that one can express the bulk field in terms of the associated CFT operator via 
\bea
\Phi^{{}^{\rm KLL}}_h(X) = \int \! d^2x\spc K(X; x) \spc {\cal O}_h(x),
\eea 
where  $K(X;x)$ denotes a suitable smearing function, that solves the wave equation in the bulk.
This is the Kabat, Lifschytz and Lowe prescription \cite{bulk2}. 

The KLL prescription has several short-comings. The map presumes the existence of a gravity dual: rather than reconstructing the extra dimensional physics, it makes explicit use of the classical bulk geometry.
Moreover, since the kernel $K(X;x)$ depends on the geometry,  $\Phi^{{}^{\rm KLL}}_h(X)$ is a state dependent operator. Finally, there appears to be an obstruction to the existence of a well-defined smearing function for black hole  space times \cite{bulk3}. 

Given these issues, it would clearly be desirable to find a definition of local bulk fields, that is  (i) inherent to the CFT, (ii) state-independent, and (iii) applicable to black hole space-times. In this note, we will propose such an intrinsic CFT definition for AdS${}_3$/CFT${}_2$. 

Our proposal makes use of Ishibashi boundary states \cite{ishi}, twisted via a cross cap identification. Let $| h\rangle$ denote the primary state with (equal left and right) conformal weight $h$: 
$L_0|h\rangle = \bar{L}_0| h \rangle = h | h \rangle$. Algebraically, the cross cap state $| \! | h\rangle \! \nspc\rangle_{\nspc \rm \otimes}$ is defined as the unique state spanned by descendents of  $| h\rangle$~such~that
\bea
\label{ishi}
\Bigl(L_{-n} - (-1)^n \bar{L}_n\Bigr) \li \! \li h\ra \! \nspc\ra_{\! \otimes} = 0.
\eea
Geometrically, the twisted boundary  state $| \! |h\rangle \! \nspc\rangle_{\nspc \otimes}$ cuts a hole in the surface on which the CFT lives, identifies diametric opposite points on the edge of the hole, and projects onto the Virasoro representation labeled by  $h$. 

The state-operator map associates to the  state $| \! |h\rangle \! \nspc\rangle_{\nspc \otimes}$, viewed as obtained via radial quantization, a local operator $\Phi_h(0,y)$ through the relation
\bea
\label{phicstate}
\Phi_\hh(0,y) \li 0 \ra = y^{L_0+ \bar{L}_0}\, \li\! \li h \ra\! \nspc \ra_{\! \otimes}.
\eea
Here $y$ is a scale modulus introduced by the boundary state. Indeed, adding a cross-cap decreases the Euler number of the surface by one, and thus adds three real shape parameters, which we can think of as the location (taken to be the origin in (\ref{phicstate})) and the size of the hole. By moving the origin to some arbitrary location $(z,\bar{z})$, we thus obtain an operator $\Phi_h(z,\bar{z},y)$ defined on a 3-dimensional space. This is our proposed CFT definition of the bulk operator associated to the local operator ${\cal O}_h(z,\bar{z})$. 
Note that the state (\ref{phicstate}) is normalizable as long as $y<1$.

\begin{center}
{ \bf Testing the proposal}
\end{center}

\vspace{-2mm}

 The bulk field $\Phi_\hh(z_0,\bar{z}_0,y)$ cuts out a circular hole~of radius $y$ centered around the point $(z_0,\bar{z}_0)$, while
 gluing together diametric opposite points via the identification\footnote{In an earlier version of this note, the orientation flip $(-1)^n$ was omitted. We thank Tadashi Takayanagi for explaining the need for this sign flip in the closely related proposal for local bulk operators put forward in \cite{tadashi-etal}. For an elegant geometrical interpretation for the extra $(-1)^n$ factor, see also \cite{hirosi-yu}.}
\bea
\label{circle}
{\bar{z} - \bar{z}_0} = -\frac{{y^2}}{z-z_0},
\eea
We can write this relation as a global $SL(2,\mathbb{R})$  transformation $\bar{z} = \frac{a z + b}{c z+d}$
%\footnotesize \begin{array}{c}
with ${y = 1/c}, { z_0 = -d/c}$ and ${\bar{z}_0 = {a}/{c} } $. The bulk field thus naturally lives on an $SL(2,\mathbb{R})$ group manifold, or a subspace thereof.  In the following, we will often denote the bulk coordinate $X$ by the group element $g =${\footnotesize $ \Bigl(\begin{array}{cc}\! a\! &\! b\! \\[-.5mm] \! c\! &\! d\! \end{array}\Bigr)$}. %, and write~$\Phi_h(g)$.

We wish to verify that the bulk field $\Phi_h(g)$ defined by (\ref{phicstate}) has the required properties (\ref{phitoo}) and (\ref{waveeq}). Property (\ref{phitoo}) is evident, since $P_0 = \lim_{y\to 0 } y^{L_0 + \bar{L}_0 - 2h}$ is a projection onto the primary state $|h \rangle = {\cal O}_h(0) | 0\rangle.$  Property (\ref{waveeq}) is a less trivial statement, that only holds to leading order at large $N$. Indeed, since our definition (\ref{phicstate}) does not presume a free bulk theory, it should automatically incorporate all interactions and $1/N$ corrections. 

How can a state-independent operator (\ref{phicstate}) satisfy a state dependent wave equation (\ref{waveeq})?
A partial answer is that in 2+1-D gravity, and outside of any matter sources, the background geometry locally looks like AdS${}_3$. So the state dependence manifests itself in the form of non-trivial global transition functions on the variable $g$. 
\def\hhh{{}}

For concreteness, consider a matrix element of the bulk field $\Phi_h(g)$ between two highly excited states 
\bea
\label{matrixelt}
\phi_h\bigl[ \mbox{\scriptsize $\! \raisebox{.25pt}{\scriptsize $\begin{array}{cc} 1\nspc \! &\!\nspc  2 \\[-.5mm] 3\nspc \! &\! \nspc 4 \end{array}$} \! $}\bigr] (g) = \la h_3,\nspc h_4
\li \Phi_{\hh}(g) \li \spc h_1, \nspc h_2\ra\qquad\qquad \\[3mm]
\textstyle h_1\nspc \nmpc =\nspc  h_2\nspc\nmpc  = \nspc \frac 1 2  \Delta, \ \ \ h_3\nspc\nmpc = \nspc \frac 12 (\Delta\nspc\nmpc +\nspc \omega\nspc\nspc +\nspc\nspc \ell), \ \ \ h_4
\nspc\nmpc = \nspc \frac 1 2(\Delta \nspc\nmpc +\nspc  \omega\nspc\nspc - \nspc\nspc \ell)
\nonumber 
\eea
The in-state $| h_1,h_2\rangle = {\cal O}_\Delta(0)| 0 \ra$ is a primary state with conformal weight $\Delta \gg c/12$. 
In the gravity dual, it is describes a non-rotating BTZ black hole with mass $M = \Delta - \frac{c}{12} 
$. The black hole space-time is obtained from the $SL(2,\mathbb{R})$  group manifold by taking the quotient \cite{BTZ}
\bea
\label{btzbc}
g \sim g_{{\nspc }_L} g\spc g_{{\nspc }_R} , \qquad
g_{{\nspc }_L} = g_{{\nspc }_R} =
e^{\pi r_{\! +} \sigma_2} .
\eea
Here $r_+ = \sqrt{8 G_N M} = \sqrt{\frac{24\Delta}{c}-1}$ is the black hole radius.  (We use $R_{\rm AdS} = 1$ units.)
The out-state in (\ref{matrixelt}) represents the black hole with a small extra mass $\omega$ 
and angular momentum $\ell$.
It is convenient to parametrize $g=  e^{\frac{i}{2} (\hat\varphi  + \hat{t}) \sigma_2} e^{i r \sigma_1}  e^{\frac{i}2 (\hat{\varphi}  - \hat{t}) \sigma_2}$, with
$\hat\varphi \nspc =\nspc 2r_{\! +} \varphi$ and  $\hat{t} \nspc =\nspc  2r_{\! +} t$, and split off the time and angular dependence via \bea
\phi_h\bigl[ \mbox{\scriptsize $\! \raisebox{.25pt}{\scriptsize $\begin{array}{cc} 1\nspc \! &\!\nspc  2 \\[-.5mm] 3\nspc \! &\! \nspc 4 \end{array}$} \! $}\bigr] 
(g) = e^{-i\omega t} e^{i \ell \varphi} \phi_h\bigl[ \mbox{\scriptsize $\! \raisebox{.25pt}{\scriptsize $\begin{array}{cc} 1\nspc \! &\!\nspc  2 \\[-.5mm] 3\nspc \! &\! \nspc 4 \end{array}$} \! $}\bigr] (r).\eea

\def\ZZ{\,\spc \mbox{\small \!\!\nspc $\spc Z$}}

The amplitude (\ref{matrixelt}) transforms in a well-prescribed way under 
 conformal transformations $(z,\bar{z}) \to (w(z),\bar{w}(\bar{z}))$. The Ishibashi state $| \! |h\rangle \! \nspc\rangle_{\nspc \otimes}$ is inert under reparametrizations that leave the location of the boundary circle (\ref{circle}) fixed.
In addition, the primary in- and out-states are annihilated by half of the Virasoro generators. This gives us the freedom to evaluate (\ref{matrixelt}) in our favorite coordinate system. 

\def\mmu{\mu}

How can we detect the BTZ monodromy (\ref{btzbc}) in the CFT? 
At large  $c$, a CFT amplitude selects a natural coordinate system, specified via the expectation value  
of the stress-energy tensor. Thanks to the anomalous transformation property of $T(z)$, one can always find local coordinates $(\ZZ,\bar{\ZZ})$ such that \bea\label{uni}
\la T(\ZZ)\ra = \la \bar{T}(\bar{\ZZ})\ra =  0.\eea We call $(Z,\bar{Z})$ the `uniformizing coordinate system'.
It associates to the amplitude a constant curvature metric 
$ds^2 =  e^\phi dz d\bar{z} = \frac{dZ d\bar{Z}}{(Z-\bar{Z})^2}$ with $\frac{c}{6} \la T\ra = -\frac{1}{2} (\partial \phi)^2 + \partial^2\phi$.

For our matrix element (\ref{matrixelt}) we have $\la T(z)\ra={ \Delta}/{z^2}$. This is uniformized by  
\bea
\label{uniformz}
\ZZ(z)= z^{ir_{\! +}} \qquad {\rm with} \qquad r_{\!+}^2 = \textstyle {\frac{24 \Delta}{c}-1}.
\eea 
The coordinates $(\ZZ,\bar{\ZZ})$ are multivalued: under a full rotation $z \to e^{2\pi i} z$, they undergo a monodromy specified by the same hyperbolic $SL(2,\mathbb{R})$ elements $g_{{\nspc }_L}$ and $g_{{\nspc }_R}$  
that characterize the BTZ geometry.  The corresponding 2-D constant curvature  metric describes a hyperbolic cylinder with two asymptotic regions corresponding to the initial and final  states.
\def\smpc{\hspace{.5pt}}
\def\Fto{{}_{\mbox{\tiny 2}}\!\smpc \spc F_{\!\spc \mbox{\tiny 1}}}

\begin{figure}[t]
\begin{center}
\includegraphics[width=0.35\textwidth]{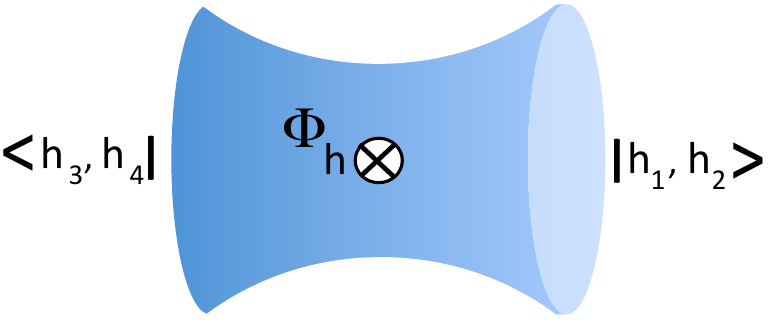}
\vspace{-2mm}
\caption{In a semi-classical  treatment, the matrix element (\ref{matrixelt}) equals the 2D Liouville action associated to a hyperpoblic cylinder with a single cross-cap. The moduli space of this hyperbolic surface is isomorphic to the BTZ black hole space-time. }
\vspace{-3mm}
\end{center}
\end{figure}
We can now use the $(\ZZ,\bar{\ZZ})$ coordinates to associate an $SL(2,\mathbb{R})$ group element $g$ to the boundary state (\ref{phicstate}), via the procedure described below eqn (\ref{circle}). When we transport $\Phi_h(g)$ around the heavy operator ${\cal O}_\Delta(0)$,  the group element $g$ does not come back to itself, but undergoes the BTZ holonomy (\ref{btzbc}). We conclude that {\it the Teichm\"uller space of the hyperbolic cylinder with a cross-cap (shown in fig 1) is isomorphic to the BTZ black hole space-time.}
This is our first piece of evidence that, at large $c$, the bulk field $\Phi_h(g)$ lives on the classical black hole background.

To provide more quantitative support for our proposal,  let us set out to compute the amplitude (\ref{matrixelt})  and compare the answer with the bulk mode 
that solves the wave equation (\ref{waveeq}) in the BTZ geometry.  This bulk mode takes the form  $f_{\omega \ell}^\hhh (t,\varphi,\rho)  
= e^{-i\omega t} e^{i \ell \varphi} f_{\omega \ell}^\hhh(\rho)$ with
\bea
\label{modefn}
%\qquad\qquad\qquad \\[2.5mm]
f_{\omega \ell}^\hhh(\rho) = \rho^{h} (1\! -\! \rho)^{\frac {i \omega}{2r_{\!\mbox{\tiny +}\!\!\nspc}}}\spc \Fto\bigl(\textstyle  \hh \! +\nspc \nspc \frac {i(\omega  + \ell)}{4r_{\! +}}, \hh \! +\nspc \frac {i(\omega  - \ell)}{4r_{\! +}}, 2\hh\spc; \rho \bigr)
\ \ \
\eea
Here $%F(a,b,c\spc ; z)=
\Fto(a,b,c\spc ; z)$ denotes the ordinary hypergeometric function and $\rho = {r_{\! +}^2}/{r^2}$ parametrizes the radial coordinate. %\footnote
{
In fact, the mode function constitutes an $SL(2,\mathbb{R})$ matrix element,  $f_{\omega \ell}^\hhh(g) =  \langle \hh, \textstyle\nspc \frac {i(\omega  + \ell)}{4r_{\! +}} |\spc g\spc | \hh, \nspc \frac{i(\omega  - \ell)}{4r_{\! +}}\rangle$,
which evidently satisfies the free bulk wave equation (\ref{waveeq}).} % of a particle of mass $m_h$. 

To compute the CFT  matrix element (\ref{matrixelt}), it is convenient to start in Euclidean signature, define $\Phi_h(g)$ as the operator that pokes a hole in the 2D Euclidean space time, and then Wick rotate back to Lorentzian signature.  Moreover, we will choose to work in the uniformizing coordinate system $(\ZZ,\bar{\ZZ})$ introduced in (\ref{uni})-(\ref{uniformz}). This choice will greatly facilitate our analysis.

A CFT amplitude on a surface  with a cross-cap is most conveniently analyzed by introducing the so-called Schottky double,
 as shown in fig 2.     In our case,  $\Sigma$ is a cylinder with a circular hole, and its Schottky double  $\tilde{\Sigma}$ is two cylinders connected via a narrow bridge. $\tilde{\Sigma}$  admits an (orientation reversing) involution that identifies diametric opposite points on the circular boundary of $\Sigma$. The reflection symmetry restricts its cross ratio $\ZZ$  to be real 
 $$
\ZZ = \bar{\ZZ} \equiv \rho.$$
A sphere with two punctures and a cross-cap has one single real modulus.

Since the boundary reflects left-moving into right-moving modes, the involution interchanges the two chiral halves of the CFT. 
Moreover, thanks to the projection onto the conformal sector $h$ in the intermediate channel,  the CFT amplitude on the double $\tilde{\Sigma}$  takes the form of a single non-chiral conformal block 
${\cal F}_{\nspc h}\!\smpc\bigl[ \mbox{\scriptsize $\! \raisebox{.25pt}{\scriptsize $\begin{array}{cc} 1\nspc \! &\!\nspc  2 \\[-.5mm] 3\nspc \! &\! \nspc 4 \end{array}$} \! $}\bigr](
\ZZ,\bar{\ZZ})$, which in turn factorizes into the product of two chiral blocks:
\bea
\label{cblock}
{\cal F}_{\nspc h}\!\smpc\bigl[ \mbox{\scriptsize $\! \raisebox{.25pt}{\scriptsize $\begin{array}{cc} 1\nspc \! &\!\nspc  2 \\[-.5mm] 3\nspc \! &\! \nspc 4 \end{array}$} \! $}\bigr](\ZZ,\bar{\ZZ}) =  \bigl|\spc \Psi_{\nspc h}\!\smpc\bigl[ \mbox{\scriptsize $\! \raisebox{.25pt}{\scriptsize $\begin{array}{cc} 1\nspc \! &\!\nspc  2 \\[-.5mm] 3\nspc \! &\! \nspc 4 \end{array}$} \! $}\bigr](\ZZ)\spc \bigr|^2.
\eea
Here we absorbed the product of OPE coefficients $C_{12h}C_{h34}$ into the normalization of the conformal block.
The $\ZZ$-dependence of the conformal blocks is universal and completely fixed by conformal invariance.

\begin{figure}[t]
\begin{center}
\includegraphics[width=0.14\textwidth]{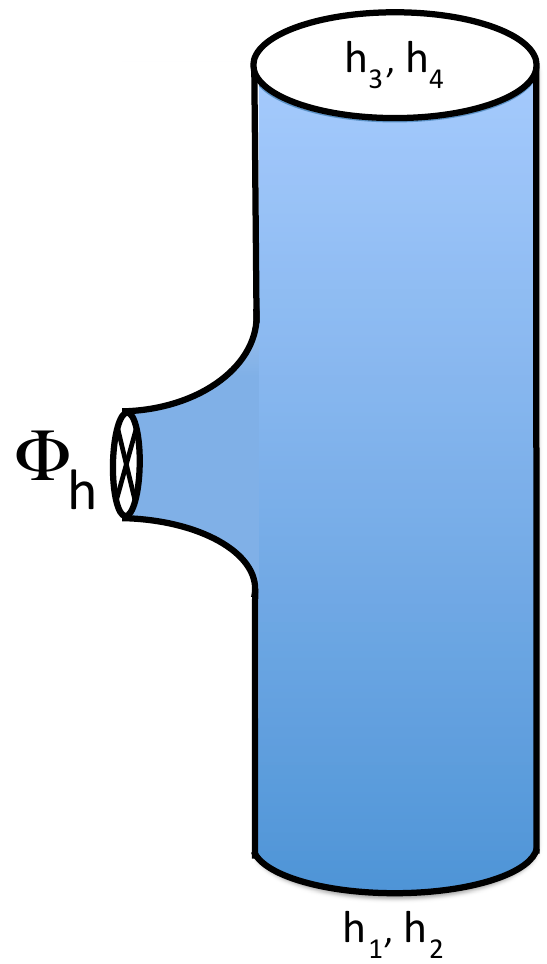}~~~~~~~~~~~
\includegraphics[width=0.225\textwidth]{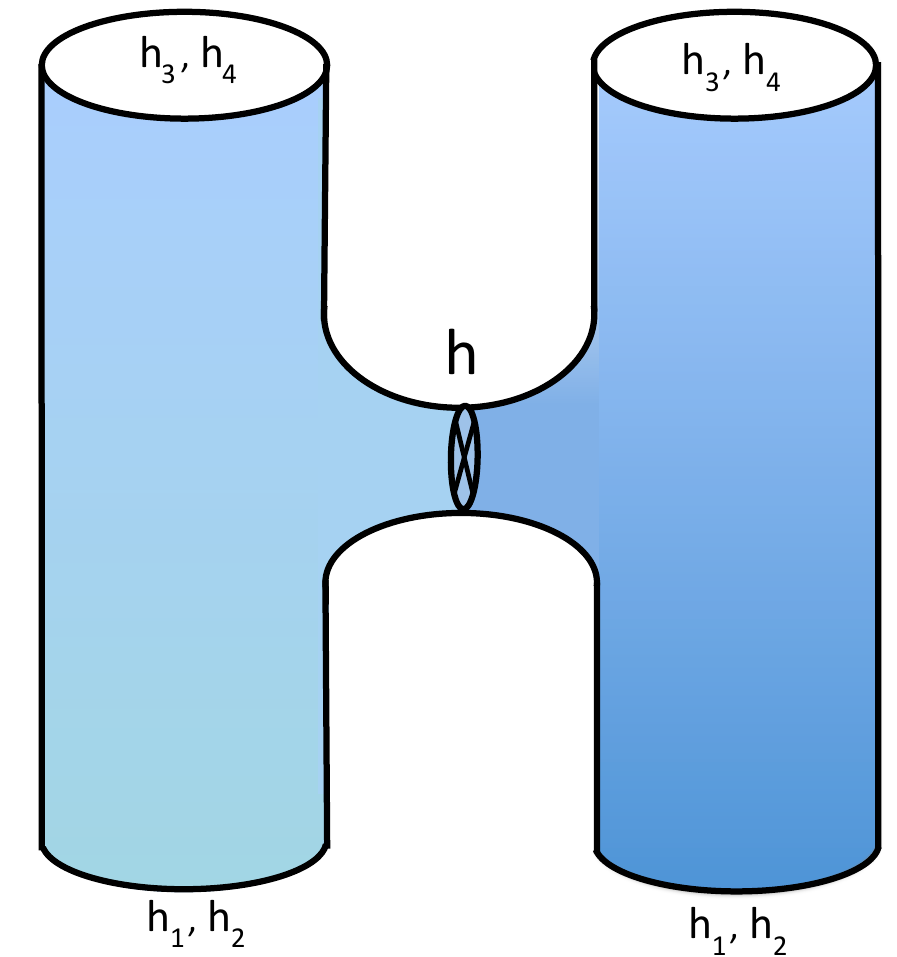}~~~~~
\end{center}
\caption{The CFT amplitude with the insertion of an Ishibashi boundary state (left) and its Schottky double (right). Due to the projection onto the conformal sector $h$ in the intermediate channel, 
the amplitude is given by a single conformal block.}
\end{figure}

\def\ccL{L}
\def\hhn{\{n\}}

The amplitude  $\phi_h\bigl[ \mbox{\scriptsize $\! \raisebox{.25pt}{\scriptsize $\begin{array}{cc} 1\nspc \! &\!\nspc  2 \\[-.5mm] 3\nspc \! &\! \nspc 4 \end{array}$} \! $}\bigr] %_{\omega_L,\omega_R}^\hhh
(g)$ is obtained by taking the square root of the amplitude on the double $\tilde{\Sigma}$. 
\bea
\label{plan}
\phi_h\bigl[ \mbox{\scriptsize $\! \raisebox{.25pt}{\scriptsize $\begin{array}{cc} 1\nspc \! &\!\nspc  2 \\[-.5mm] 3\nspc \! &\! \nspc 4 \end{array}$} \! $}\bigr] %_{\omega_L,\omega_R}^\hhh
(\rho) =\,\Bigl({\cal F}_{\nspc h}\!\smpc\bigl[ \mbox{\scriptsize $\! \raisebox{.25pt}{\scriptsize $\begin{array}{cc} 1\nspc \! &\!\nspc  2 \\[-.5mm] 3\nspc \! &\! \nspc 4 \end{array}$} \! $}\bigr] (\rho,\rho) \Bigr)^{1/2}%(z,\bar{z})_{|z=\bar{z}=\rho} 
\eea
So our task is: (i) compute the conformal block, (ii) take the square root, (iii) compare the result with the mode function (\ref{modefn}) in the BTZ black hole background ({\it c.f.} \cite{Krasnov}).

Virasoro conformal blocks are uniquely determined by the conformal Ward identity. An explicit expression is not available yet, however, though exact or semi-classical properties  are known. Known exact results are (a)~Zamolodchikov's recursion formula \cite{zam} relating Virasoro and global conformal blocks and (b) the modular `fusion' matrices, obtained by Ponsot and Teschner from Liouville CFT and quantum Teichmuller theory \cite{pt}. Semi-classical expressions have recently been obtained in \cite{fitz}. Unfortunately, none of the known results allow us to read off the specific answer that we need.

Given this state of affairs, it's a reasonable strategy at this point to invert the sequence of step (i)-(iii), and first deduce the desired expression for the 2D conformal block that we need in order to  find a precise match 
\be
\label{match}
\phi_h\bigl[ \mbox{\scriptsize $\! \raisebox{.25pt}{\scriptsize $\begin{array}{cc} 1\nspc \! &\!\nspc  2 \\[-.5mm] 3\nspc \! &\! \nspc 4 \end{array}$} \! $}\bigr](\rho)\, =\, f_{\omega\ell}^\hhh(\rho)
\ee
between the CFT amplitude and the bulk mode function. Imposing this match, we deduce that the conformal block should take the following form
  \bea
 \label{express}
{\Psi}_{\nspc h}\!\smpc\bigl[ \mbox{\scriptsize $\! \raisebox{.25pt}{\scriptsize $\begin{array}{cc} 1\nspc \! &\!\nspc  2 \\[-.5mm] 3\nspc \! &\! \nspc 4 \end{array}$} \! $}\bigr]
 (\ZZ) = \textstyle \ZZ^h\Fto\bigl(h\nspc +\nspc \frac{i}{2r_{\! + }\! }\spc h_{13}, h \nspc +\nspc  \frac{i}{2r_{\! + }\!\!}\, h_{24}, 2\hh\spc ; \ZZ \bigr)
 \eea
To see this, note that taking the chiral conjugate of the conformal block ${\Psi}_{\nspc h}\!\smpc\bigl[ \mbox{\scriptsize $\! \raisebox{.25pt}{\scriptsize $\begin{array}{cc} 1\nspc \! &\!\nspc  2 \\[-.5mm] 3\nspc \! &\! \nspc 4 \end{array}$} \! $}\bigr]
 (\ZZ)$ amounts flipping the sign of $\frac{i}{2r_{\! + }\! }\spc h_{13}$ and $\frac{i}{2r_{\! + }\! }\spc h_{24}$ inside the argument of the hypergeometric function. Then,
using the standard identity 
$
\Fto(a,b,c\spc ; \ZZ) = (1-\ZZ)^{c-a-b} \Fto(c-a,c-b,c\spc ; \ZZ)$, direct inspection shows that the CFT amplitude (\ref{cblock})-(\ref{plan}) reproduces the expression (\ref{modefn}).

Formula (\ref{express}) for the conformal blocks is a conjecture. We will now present two pieces of supporting evidence. 

The uniformizing coordinate system (\ref{uniformz}) has the special property that the semi-classical Virasoro conformal blocks effectively reduce to global conformal blocks \cite{fitz}. Consider the definition $
\Psi_{\nspc h}\!\smpc\bigl[ \mbox{\scriptsize $\! \raisebox{.25pt}{\scriptsize $\begin{array}{cc} 1\nspc \! &\!\nspc  2 \\[-.5mm] 3\nspc \! &\! \nspc 4 \end{array}$} \! $}\bigr](\ZZ) = \la {\cal O}_{4}(\infty){\cal O}_2(1) \ri {\cal P}_h \li {\cal O}_3(\ZZ) {\cal O}_1(0)\ra$
of the  chiral  block.
Here ${\cal P}_h  $ %= \, \sum \, \frac 1 {{\cal N}_{\nspc \hhn}\!\!\!\!\!\!}\;\; { {\ccL}^\dag_{\hhn} \li h \ra \la h \li {\ccL}_{\hhn}}
 is the projection operator onto the sector spanned by all descendents  %$\ccL^\dag_{\hhn}\li h\ra = \ccL_{-n_1}^{k_1}%\ccL_{-n_2}^{k_2}...\ccL_{-n_s}^{k_s}\li h\ra$  
 of $\li h\ra$. From the commutator $[\ccL_{n},\ccL_{-n}] = 2n \spc \ccL_0 + \frac{c}{12} n (n^2-1)$ we read off that the norm 
of the descendent states grows linear with $c$, except for the `global descendents' of the form  $\ccL_{-1}^k\li h \ra$.
Due to the special choice of the coordinate system (\ref{uniformz}),  the $L_{-n}$ operators do not produce any other large factors proportional to the conformal weight $\Delta$. So, as argued in \cite{fitz},  we may replace the projection operator ${\cal P}_k$ by a restricted sum
over global descendents only. 

 \def\is{\! & \! = \! & \!}

Global conformal blocks are known to satisfy a differential equation of the form
${\ccL}_{\rm tot}^2
 {\cal F}
 _{\nspc h}\!\smpc 
 (z,\bar{z})\spc =\spc 2 m^2_h\spc
{\cal F}
_{\nspc h}
 (z,\bar{z})$
 with ${\ccL}_{\rm tot}^2 = L^2 + \bar{L}^2$ the Casimir of the global conformal algebra $\mathfrak{so}(2,2)$  acting on the intermediate channel. The solution factorizes into  chiral global conformal blocks
 given in terms of the hypergeometric function via  $\Psi_h = z^h \Fto\bigl(h\nspc +\nspc h_{13}, h \nspc + \nspc h_{24}, 2\hh;  z\bigr)$,
 with $h_{ij} = h_i \nspc - h_j$.  In our setting, we also need to take into account that in the $(\ZZ,\bar{\ZZ})$ system, the generator of scale transformations ${}\ZZ\frac{\partial\ }{\partial Z} = \frac{i}{2r_+\!\!} \, z\frac{\partial\ }{\partial z}$ is rescaled  by a factor $\frac{i}{2r_+\!\!}\,$ relative to the standard $L_0$-generator.
 This renormalizes $h_{ij}$ to $\frac{i}{2r_+\!\!\!} \, h_{ij}$. This  is precisely what we need to recover formula (\ref{express}).

Formula (\ref{express}) receives additional support from known results in Liouville conformal field theory \cite{pt}, or equivalently, from the identification \cite{HV} between Virasoro conformal blocks and the Hilbert states obtained by quantizing  Teichm\"uller space \cite{nt} \cite{pt}. 
The central result of \cite{pt} is that the monodromy properties of Virasoro blocks are identical to those of the invariant tensors obtained by gluing together two Clebsch-Gordan coefficients of the quantum group $U_q(\mathfrak{sl}(2))$. Using this abstract tensor categorical definition of conformal blocks, it was found that the 4-point conformal blocks have (in a suitable normalization) an algebraic representation in terms of b-deformed hypergeometric functions
\bea
\label{pto}
\Psi^b_{h}\!\smpc \mbox{\small $\bigl[\! \raisebox{.25pt}{\scriptsize $\begin{array}{cc} 1\nspc \! &\!\nspc  2 \\[-.25mm] 3\nspc \! &\!\nspc  4 \end{array}$} \! \bigr]$}(x)\nspc = e^{2\pi \alpha_h x}
F_{b}\bigl(\alpha_h\! + \nspc
\alpha_{13}
, \alpha_h\!
+\nspc  \alpha_{24}
, 2\alpha_h ;\nspc -ix\bigr)\ \
\eea
where  $\alpha_i$ are the Liouville momenta $h_i = \alpha_i(Q\nmpc -\nmpc \alpha_i)$, and $q$, $b$ and $Q$ are related to $c$ via $q = e^{i\pi b^2}$, $c = 1+6 Q^2$ and $Q=b+b^{-1}$. The variable $\ZZ_b = -e^{2\pi b x}$ defines a quantum coordinate $\ZZ_b$, on which $U_q(\mathfrak{sl}(2))$ acts via suitable $q$-deformed $SL(2,\mathbb{R})$  transformations.  

The expression (\ref{pto}) is an eigen function of the q-deformed %$U_q(\mathfrak{sl}(2))$ 
Casimir operator $C_{12}$ acting on the intermediate channel \cite{pt}.
Hence, in the small $b$ =  large~$c$ limit, it is reasonable to identify the quantum group coordinate $Z_b$ with our uniformizing coordinate $Z$. We further have
\bea
 \alpha   \simeq b \spc h \qquad 
{\alpha_{13}}   \simeq  \, \frac {ib} {2r_{\! + }\!}\spc {h}_{13} 
, \qquad
 {\alpha_{24}}    \simeq \, \frac{i b}{2r_{\! +}\!} \spc {h}_{24} 
 \nonumber
\eea
and the b-deformed hypergeometric function reduces to 
\bea
F_b(\alpha,\beta,\gamma ;-i x)\ \  \raisebox{-3pt}{${\longrightarrow}\atop{\mbox{\scriptsize{$b\!\nspc \to\!\nspc 0$}}}$}\  \ \Fto(\aa,\bb,\cc\spc; \ZZ)\nonumber\\[1mm]
\alpha  = b \aa, \ \ \beta  = b \bb, \ \ \gamma  = \spc b\spc \cc .\qquad \nonumber
\eea
So in the limit of large central charge $c$ we again recover the desired result~(\ref{express}).
%This match is our main piece of supporting evidence for the proposed interpretation of $\Phi_h(g)$ as a local bulk field.

 %with  $\mu_{ij} \! = {\Delta_{ij}}/{2r_+}$. 
\medskip

\begin{center}
{\bf 2-point function} 
\end{center}

\vspace{-2mm}

What about the higher point functions? Consider the 2-point amplitude between two excited states 
\bea
G_h(g_1,g_2) = \la \spc \Delta \spc \li \Phi_{\hh}(g_1)\spc \Phi_h(g_2) \li \spc \Delta \spc \ra,
\label{green}
\eea
of equal conformal weight $\Delta \gg \frac{c}{12}$. We will assume that the bulk locations $g_1$ and $g_2$  are slightly smeared, so that the highest frequencies that contribute to the 2-point function remain sub-Planckian. In this regime, the 2-point function is expected to reduce to the Hartle-Hawking propagator of a massive scalar field of mass $m_h$ in the BTZ black hole background. On the CFT side, this result arises as follows.

Inserting a complete set of states factorizes the 2-point function into a product of 1-point matrix elements
\bea
\quad \sum_{h_3,h_4}\, \la \spc \Delta \spc \li \Phi_{\hh}(g_1)\li h_3,h_4\ra\la h_3,h_4\ri \Phi_h(g_2) \li \spc  \Delta \spc \ra %+ \ldots
\label{sum}
\eea
By our low energy assumption, the contribution of the descendent states is subleading in $1/N$. Using our result for the matrix elements of $\Phi_h(g)$ between energy-momentum eigenstates, we find that (after analytic continuation to Minkowski signature) the 2-point function takes the form
\bea
\sum_{\ell} \int_0^\infty \!\!\!\!\! d\omega \,  \Bigl( n_{{}_+}\!(\omega)\spc \spc  f^*_{\omega \ell}(g_1)\spc f^\hhh_{\omega\ell}(g_2) + n_{{}_-}\!(\omega)\spc f^\hhh_{\omega \ell}(g_1)\spc  f^*_{\omega\ell}(g_2)\Bigr)\nonumber
\eea
where $n_{{}_\pm}\!(\omega)$ denote the  level densities of intermediate  states that contribute in the sum (\ref{sum}).  

The form of $n_{{}_\pm}\!(\omega)$  is determined by taking the limit 
(\ref{phitoo}) where $\Phi_h(g_1)$ and $\Phi(g_2)$ both approach the AdS boundary, where they reduce to local CFT operators. In this  limit, it has already been shown \cite{fitzone} that the 2-point function (\ref{green}) receives its dominant contribution from the identity conformal block and reproduces the boundary-to-boundary propagator in the black hole background. This result can be viewed as a confirmation of the eigenvalue thermalization hypothesis (ETH): the 2-point function (\ref{green}) for $g_1$ and $g_2$ close to the boundary behaves as the CFT 2-point function ${\rm tr}\bigl(\rho_\beta {\cal O}_h(x_1) {\cal O}_h(x_2)\bigr)$ in the thermal state $\rho_\beta = e^{-\beta H}$ with temperature $\beta = \frac{2\pi}{r_+}$ equal to a BTZ black hole of mass $M = \Delta - \frac {c}{12}$. This shows that the spectral densities $n_\pm(\omega)$ take the form of thermal probability distributions with 
\bea 
\label{thermal}
n_+(\omega) = e^{-\beta \omega} n_-(\omega).
\eea
Via the found match with the mode functions, we can now extend this result to the bulk and confirm that
(in the low energy regime in which the expansion (\ref{sum}) is valid) the bulk 2-point function (\ref{green}) coincides with the Hartle-Hawking propagator.

\begin{center}
{\bf Concluding comments} 
\end{center}

\vspace{-2mm}

Our results have some bearing on the firewall puzzle. After analytic continuation to Lorentzian signature, the individual mode functions $f^\hhh_{\omega\ell}(g)$
exhibit a $(\rho-1)^{i \omega/r_+}$ branch-cut at the location of the black hole horizon. This behavior is as expected, given that the mode functions  carry a definite
energy as seen from outside. However, since the  spectral densities $n_\pm(\omega)$ satisfy (\ref{thermal}), the 2-point function exhibits perfectly smooth behavior at the black hole horizon. This is also no surprise, since we're simply reversing Hawking's original derivation and  its arrow of implication. Indeed, it is natural to propose that the two basic characteristics of a black hole imply each other
$$
\begin{array}{c}{\mbox{Smoothness of} }\\ {\mbox{the event horizon} }\end{array} \ \ \ %{\textcolor{black}{\Longrightarrow} \atop \textcolor{black}{\Longleftarrow}} 
\Longleftrightarrow \ \  \begin{array}{c}{\mbox{Thermality $\&$ analyticity}}\\ {\mbox{of the 2-point function}}\end{array}
$$

Analyticity of the 2-point function is a natural consequence of our geometric CFT definition of bulk fields. Mode functions are identified with  conformal blocks, which are analytic functions of their arguments. Thermality is a manifestation of the ETH, which asserts that energy eigenstates behave like thermal states for few point functions that only probe a small subsector of the total quantum system. The two properties combined appear sufficient to conclude that the 2-point function (\ref{green}) is smooth across the horizon. 

There is some fine-print, however. In the expansion (\ref{sum}), we omitted the contribution of descendent states.  In space-time language, this presumes that the field operator $\Phi_h(g)$  produces only low energy modes, and does not excite any boundary gravitons. So our bulk fields have the properties of low energy effective quantum fields provided they are restricted to act within a `code subspace' of the total CFT Hilbert space, spanned by states that are compatible with a given semi-classical geometry.

Finally, it should be noted that the definition (\ref{phicstate}) of the bulk field $\Phi(y,z,\bar{z})$ is in fact less unique that it appears. Whereas Ishibashi states $|\! \spc | h \rangle\! \rangle_{\nspc \otimes}$ are invariant under reparametrizations that leave the location of the circle (\ref{circle}) fixed, it is not invariant under reparametrizations that deform the location of the circle. For every closed curve $C$ surrounding a point $x$, one can find a coordinate system for which $C$ looks like a circle centered around $x$. So Ishibashi states and our bulk operators should in fact be labeled by arbitrary closed curves $C$.
The matrix element $\langle \Delta+ \omega| \Phi_h(C)|\Delta\rangle$ between some given initial and final primary state,  however, only depends on the 3 moduli $(y,z,\bar{z})$ associated with the hole created by $\Phi_h(C)$. Most of the shape parameters of $C$ can be removed by using that the initial and final states are Virasoro highest weight states. So a bulk space-time location is {\it neither} uniquely specified by the operator  $\Phi_h(C)$ {\it nor} state-independent: it is determined by the {\it relation} between the curve $C$ of the bulk operator, and the choice of initial and final state.

This redundancy is likely to play a key role in establishing an effective form of bulk locality, which dictates that space-like separated  bulk fields should, to a high degree of accuracy, commute with each other. In particular, a bulk field $\Phi_h(X)$ should commute with any given local operator ${\cal O}(x)$ on the boundary. The argumentation here mirrors the `secret sharing protocol' put forward in \cite{danetal}: since for a given initial and final state, there are many curves $C$ that map to the space-time point $X$, it should always be possible to choose a representative curve $C$ that is space-like separated from some given $x$. With this choice, locality of the boundary theory ensures that $\Phi_h(C)$ commutes with ${\cal O}(x)$.

%\begin{figure}[hbtp]
%\begin{center}
%\includegraphics[width=0.215\textwidth]{fourpointsb.pdf}~~~
%\includegraphics[width=0.248\textwidth]{pairpoints.pdf}
%\end{center}
%\end{figure}

\medskip

\begin{center}
{\bf Acknowledgements}
\end{center}
\vspace{-2mm}

We thank Daniel Harlow, Juan Maldacena, Eric Perlmutter, David Poland, Massimo Porrati, Tadashi Takayanagi and Erik Verlinde for helpful discussions.
This work  is supported by NSF grant PHY-1314198.

\end{document}